\documentclass[
 reprint,
 amsmath,
 aps,
]{revtex4-2}
\usepackage{overpic}
\usepackage{multirow}

\begin{document}
\title{Modeling of three-dimensional betatron oscillation and radiation reaction in plasma accelerators}

\author{Yulong Liu}
\author{Ming Zeng}
 \email[Corresponding author: ]{zengming@ihep.ac.cn}
\affiliation{Institute of High Energy Physics, Chinese Academy of Sciences, Beijing 100049, China}
\affiliation{University of Chinese Academy of Sciences, Beijing 100049, China}

\date{\today}

\begin{abstract}
Betatron oscillation is a commonly known phenomenon in laser or beam driven plasma wakefield accelerators. In the conventional model, the plasma wake provides a linear focusing force to a relativistic electron, and the electron oscillates in one transverse direction with the betatron frequency proportional to $1/\sqrt{\gamma}$, where $\gamma$ is the Lorentz factor of the electron. In this work, we extend this model to three-dimensional by considering the oscillation in two transverse and one longitudinal directions. The long-term equations, with motion in the betatron time scale averaged out, are obtained and compared with the original equations by numerical methods. In addition to the longitudinal and transverse damping due to radiation reaction which has been found before, we show phenomena including the longitudinal phase drift, betatron phase shift and betatron polarization change based on our long-term equations. This work can be highly valuable for future plasma based high-energy accelerators and colliders.
\end{abstract}

\maketitle

\section{Introduction}
The new generation of accelerators, using plasma as the acceleration media, offer high acceleration gradient in the order of 10–100 $\rm GV/m$ and strong transverse focusing field~\cite{TTajimaPRL1979,PChenPRL1985,Blumenfeld2007nature}. Depends on the driver type, the plasma accelerators are named laser wakefield accelerators (LWFAs), which are driven by laser pulses, and plasma wakefield accelerators (PWFAs), which are driven by charged particle beams. When a high intensity laser pulse ($\gtrsim 10^{18} \ \rm W/cm^2$) or a high current particle beam ($\gtrsim 1\ \rm kA$) propagates through an underdense plasma,  the radiation pressure of the laser or the space charge of the beam expels all plasma electrons away from axis radially, leaves behind a nearly uniform ion channel. This high-intensity three-dimensional (3D) regime has been referred to as the blowout regime~\cite{Rosenzweig1991PRA, Pukhov2002APB}. In this regime, the expelled electrons are pulled back by the ion channel and thereby bubble-like plasma wake wave is created. The wake consists of a longitudinal electric field that is a function of distance behind the driver, and transversal electromagnetic fields that are proportional to the off-axis distance. Consequently, in addition to the longitudinal acceleration / deceleration, the electrons reside in the wake also perform radial oscillation, called betatron oscillation (BO), under the action of transverse focusing field, with the frequency $\omega_\beta = \omega_p\kappa/\sqrt{\gamma}$, where $\omega_p$ is the plasma frequency, $\gamma$ is the relativistic factor of the electron, and $\kappa$ is the focusing constant which takes $1/\sqrt{2}$ for the blow-out regime~\cite{KostyukovPoP2004, LuWPoP2006}.

Electrons emit synchrotron radiation when performing BO~\cite{EsareyPRE2002,RoussePRL2004,CordeRMP2013}, which affects the electrons in return. Such effect is called the radiation reaction (RR) and its classical expression is the Lorentz–Abraham–Dirac (LAD) equation or the Landau-Lifshitz equation~\cite{Dirac1938, Landaubook1975}. Because the RR force is proportional to the classical electron radius $r_e \approx 2.81 \times 10^{-15} \ \rm m$, it is generally negligible unless under extreme conditions~\cite{ColePRX2018, PoderPRX2018} or for sufficiently long interaction time~\cite{NielsenNJP2021}. The BO in a plasma accelerator is another good case for such long interaction time. The radiation leads to the energy loss of electrons and in return affects the energy-dependent betatron frequency, as well as the other beam properties, such as the energy spread and emittance~\cite{PMichelPRE2006, KostyukovJETP2006, SchroederPRSTAB2010, DengahPRSTAB2012, IYKostyukovPRSTAB2012, AFerraPousaSR2019, MZengNJP2021, GolovanovNJP2022}.

Although there are many established theories on the long-term RR damping effect of BO, their models assert the electron only moves in one plane, spanned by the longitudinal direction and one transverse direction, thus only linear polarization is considered. Moreover, these models usually neglect the longitudinal and energy oscillations during one betatron period. In this paper, we establish a 3D BO model with RR effect, which generalizes the betatron polarization from linear to elliptical, and considers both the longitudinal and energy oscillations. Long-term equations (LTEs), without resolving the betatron period, are derived and verified by numerical methods. The LTEs reproduce the previous results, such as longitudinal and transverse damping due to RR, as the special cases, and meanwhile reveal new phenomena such as betatron phase shift and polarization change.

The rest of this paper is organized as follows. Sec.~\ref{sec:EOM} gives the original form of the force, and shows the equations of motion expressed by the transverse motion only. Sec.~\ref{sec:LTE} derives the LTEs by averaging the equations of motion through one betatron period. Sec.~\ref{sec:two_regimes} discusses the different phenomena in the RR dominant regime and the betatron phase shift dominant regime. Sec.~\ref{sec:num} numerically verifies the LTEs by comparing with the code PTracker which solves the equations with the original form of force. Before start, it is worth noting that we use plasma normalization units described in Appx.~\ref{app:norm}, and some symbols and calculation rules used often during the derivation are described in Appx.~\ref{app:rules}.

\section{\label{sec:EOM}The electromagnetic field and the equations of motion}
Consider an electron with $\gamma \gg 1$ is trapped in a plasma wakefield with the longitudinal co-moving coordinate $\zeta = z-\beta_w t$, where the wake is moving in the $z+$ direction with the phase velocity $\beta_w$. Neglect the interaction between the beam particles, the electromagnetic field provided by the wake can be modeled as~\cite{LuWPoP2006}
\begin{eqnarray}
    E_z &=& E_{z0} + \lambda \zeta_1, \label{eq:Ez}\\
    \vec{E}_{\perp} &=& \kappa^2 \left(1-\lambda\right) \vec{r}, \label{eq:Er}\\
    B_{\theta} &=& -\kappa^2 \lambda r,
\end{eqnarray}
where $E_{z0}=\left.E_z\right|_{\zeta=\left<\zeta\right>}$, $\lambda=\left.d E_z/d \zeta \right|_{\zeta=\left<\zeta\right>}$, and $\vec{r}=\left(x, y\right)$ is the transverse offset. The force can be expressed as
\begin{eqnarray}
    f_z &=& -E_{z0} - \lambda \zeta_{1} + \kappa^2 \lambda (x\beta_x + y \beta_y) + f_z^{\rm rad}, \label{eq:fz}\\
    f_x &=& -\kappa^2 \left(1 -\lambda + \lambda\beta_z\right) x + f_x^{\rm rad},\\
    f_y &=& -\kappa^2 \left(1 -\lambda + \lambda\beta_z\right) y + f_y^{\rm rad}, \label{eq:fy}
\end{eqnarray}
where $\beta_x=\dot{x}$, $\beta_y=\dot{y}$, $\beta_z =\dot{z} = \beta_{z0} + \dot{\zeta_1}$, $\beta_{z0}=\beta_w + \dot{\left<\zeta\right>} = \left<\beta_z\right>$, and $\vec{f}^{\rm rad}$ is the RR force. The formulas of 3D BO with RR can be written in the form of transverse terms only (see Appx.~\ref{app:Eq_motion})
\begin{widetext}
\begin{eqnarray}
    \dot{\gamma} &=& -E_{z0}\beta_{z0} + \left(\frac{\lambda\beta_{z0}}{4}+\kappa^2\lambda-\kappa^2\right) \left(x \beta_x + y \beta_y\right) - \frac{2}{3}r_e\gamma^2\kappa^4\left(x^2 + y^2 \right), \label{eq:dot_gamma_rr}\\
    \dot{p_z} &=& -E_{z0} + \lambda\left(\frac{1}{4} + \kappa^2\right) \left(x \beta_x + y \beta_y\right) - \frac{2}{3}r_e\gamma^2\kappa^4\left(x^2 + y^2 \right),\\
    \dot{p_x} &=& -\kappa^2x + \frac{\kappa^2 \lambda}{2}\left(\left<\gamma \right>^{-2} + \beta_x^2 + \beta_y^2 \right)x - \frac{2}{3}r_e\gamma^2\kappa^4\left(x^2 + y^2 \right)\beta_x, \label{eq:dot_px_rr} \\
    \dot{p_y} &=& -\kappa^2 y + \frac{\kappa^2 \lambda}{2}\left(\left<\gamma \right>^{-2} + \beta_x^2 + \beta_y^2 \right)y - \frac{2}{3}r_e\gamma^2\kappa^4\left(x^2 + y^2 \right)\beta_y, \label{eq:dot_py_rr}
\end{eqnarray}
\end{widetext}
where
\begin{eqnarray}
    \vec{p} &=& \gamma \vec{\beta}, \label{eq:gamma_p}\\
    \beta_{z0} &=& 1 - \frac{1}{2}\left(\left<\gamma\right>^{-2} + \left<\beta_{x}^2\right> + \left<\beta_{y}^2\right> \right), \label{eq:betaz0}
\end{eqnarray}
and $r_e$ is also normalized to $k_p^{-1}$. One may note the second terms in Eqs.~(\ref{eq:dot_gamma_rr}) - (\ref{eq:dot_py_rr}), which come from the oscillation of $\beta_z$ and the modulation of $\gamma$ due to transverse oscillation, were neglected in previous works. In the following sections we show these terms lead to new phenomena.

\section{\label{sec:LTE}The long-term equations of 3D betatron oscillation}
In this section we use the same averaging method as Ref.~\cite{IYKostyukovPRSTAB2012}. We firstly introduce two complex variables
\begin{eqnarray}
    U &=& \left(x - i \kappa^{-1} \gamma^{\frac{1}{2}} \beta_x \right)e^{-i\varphi}, \label{eq:U}\\
    V &=& \left(y - i \kappa^{-1} \gamma^{\frac{1}{2}} \beta_y \right)e^{-i\varphi},
\end{eqnarray}
where
\begin{equation}
    \varphi = \int \omega_\beta dt = \kappa \int \gamma^{-\frac{1}{2}} dt
    \label{eq:varphi}
\end{equation}
is the betatron phase. Obviously $\left|U_1\right|\ll \left|\left<U\right>\right|$ and $\left|V_1\right|\ll \left|\left<V\right>\right|$ are satisfied, and we apply the rules in Appx.~\ref{app:rules} often in the following. Because the equations for $x$ and $y$ directions are symmetric, we may derive for $x$ direction only, then exchange $x$ and $y$, $U$ and $V$ for the $y$ direction. With the help of Eqs.~(\ref{eq:dot_gamma_rr}), (\ref{eq:dot_px_rr}) and (\ref{eq:gamma_p}), we may write the time derivative of Eq.~(\ref{eq:U}) as
\begin{widetext}
\begin{equation}
\begin{aligned}
    \dot{U} = &-i \frac{1}{2} \kappa^{-1} \gamma^{-\frac{1}{2}} E_{z0} \beta_{z0} \beta_x e^{-i\varphi} + i \frac{1}{3} r_e \kappa^3 \gamma^{\frac{3}{2}} \left(x^2 + y^2 \right) \beta_x e^{-i\varphi} \\
    &+ i \frac{1}{2} \kappa^{-1} \gamma^{-\frac{1}{2}} \left[\frac{\lambda \beta_{z0}}{4} + \kappa^2 \left(\lambda - 1 \right) \right] \left(x \beta_x + y \beta_y \right) \beta_x e^{-i\varphi} 
    - i \frac{1}{2} \kappa \lambda \gamma^{-\frac{1}{2}} \left( \left< \gamma \right>^{-2} + \beta_x^2 + \beta_y^2 \right) x e^{-i\varphi}. \label{eq:dot_U}
\end{aligned}
\end{equation}
\end{widetext}

In the following, we omit $\left<\right>$ on $U$ and $V$ for convenience, so that all $U$ and $V$ actually mean $\left<U\right>$ and $\left<V\right>$. We perform average on Eq.~(\ref{eq:dot_U}) to obtain (note only the terms with $e^{i0\varphi}$ survive after averaging)
\begin{widetext}
\begin{equation}
\begin{aligned}
    \dot{U} = &\frac{1}{4} E_{z0} \beta_{z0} \left< \gamma \right>^{-1} U - \frac{1}{24} r_e \kappa^4 \left< \gamma \right> \left(\left| U \right|^2 U + 2\left| V \right|^2 U - V^2 U^{*} \right) + i \frac{1}{64} \kappa \lambda \beta_{z0} \left< \gamma \right>^{-\frac{3}{2}} \left(\left| U \right|^2 U + V^2 U^{*} \right) \\
    &- i \frac{1}{16} \kappa^3 \left< \gamma \right>^{-\frac{3}{2}} \left[\left( \left| U \right|^2 + 2 \lambda \left| V \right|^2 \right) U - \left(2 \lambda -1 \right) V^2 U^{*} \right] - i \frac{1}{4} \kappa \lambda \left< \gamma \right>^{-\frac{5}{2}}U. \label{eq:dot_U_ave}
\end{aligned}
\end{equation}
\end{widetext}
By asserting $V=0$ and omitting the last three terms in Eq.~(\ref{eq:dot_U_ave}), which comes from the second terms in Eqs.~(\ref{eq:dot_gamma_rr}) - (\ref{eq:dot_py_rr}), we can reproduce Eq.~(19) in Ref.~\cite{IYKostyukovPRSTAB2012}.

The average of Eq.~(\ref{eq:dot_gamma_rr}) leads to
\begin{equation}
    \left<\dot{\gamma} \right> = - E_{z0} \beta_{z0} - \frac{1}{3} r_e \kappa^4 \left< \gamma \right>^2 \left(\left| U \right|^2 + \left| V \right|^2 \right), \label{eq:dot_gamma_ave}
\end{equation}
with the second term reproduces Eq.~(B2) in Ref.~\cite{SchroederPRSTAB2010}. $E_{z0}$ is a function of $\left<\zeta\right>$, which obeys
\begin{equation}
    \dot{\left<\zeta \right>} = \frac{1}{2} \gamma_w^{-2} - \frac{1}{2} \left<\gamma\right>^{-2} - \frac{1}{4} \kappa^2 \left< \gamma \right>^{-1} \left(\left| U \right|^2 + \left| V \right|^2 \right), \label{eq:dot_zeta_ave}
\end{equation}
where $\gamma_w = \left(1-\beta_w^2\right)^{-1/2}$ and we have used Eq.~(\ref{eq:gammaz0}).

The above averaged equations Eqs.~(\ref{eq:dot_U_ave}), (\ref{eq:dot_gamma_ave}) and (\ref{eq:dot_zeta_ave}) are already enough to predict the long-term behavior of BO. However, the equations for the complex variables are not explicit. To make them more physically meaningful, we introduce
\begin{eqnarray}
    U = \left| U \right| e^{i\Phi_x}, \\
    V = \left| V \right| e^{i\Phi_y}, \\
    \Delta \Phi = \Phi_y - \Phi_x.
\end{eqnarray}
$\left| U \right|$ has the meaning of the BO amplitude in the $x$ direction, and $\Phi_x$ the phase shift. For the $y$ direction they are similar. Thus $\Delta \Phi$ is the phase difference of the two directions. By Applying $d\left|U\right|/dt = \left(\dot{U}U^*+U\dot{U}^*\right)/2\left|U\right|$ and $\dot{\Phi}_x = \left(\dot{U} U^{*} - \dot{U}^{*} U\right)/2i \left| U \right|^2$ we get
\begin{widetext}
\begin{equation}
\begin{aligned}
    \frac{d{\left| U \right|}}{d t} = &\frac{1}{4} E_{z0} \beta_{z0} \left< \gamma \right>^{-1} \left| U \right| - \frac{1}{24} r_e \kappa^4 \left< \gamma \right> \left[\left| U \right|^3 + \left| V \right|^2 \left| U \right| \left(2 - \cos 2 \Delta \Phi \right) \right]\\
    &- \frac{1}{16}\kappa \left[ \frac{1}{4} \lambda \beta_{z0} - \kappa^2 \left(1-2\lambda\right) \right] \left< \gamma \right>^{-\frac{3}{2}} \left| V \right|^2 \left| U \right| \sin 2 \Delta \Phi, \label{eq:dot_U_ave_2}
\end{aligned}
\end{equation}
\begin{equation}
\begin{aligned}
    \dot{\Phi}_x = &\frac{1}{24} r_e \kappa^4 \left< \gamma \right> \left| V \right|^2 \sin 2 \Delta \Phi + \frac{1}{64} \kappa \lambda \beta_{z0} \left< \gamma \right>^{-\frac{3}{2}} \left[\left| U \right|^2 + \left| V \right|^2 \cos 2 \Delta \Phi \right] \\
    &- \frac{1}{16} \kappa^3 \left< \gamma \right>^{-\frac{3}{2}} \left[\left| U \right|^2 + 2 \lambda \left| V \right|^2 + \left(1-2\lambda\right) \left|V\right|^2 \cos 2 \Delta \Phi \right] - \frac{1}{4} \kappa \lambda \left< \gamma \right>^{-\frac{5}{2}}, \label{eq:dot_Phix}
\end{aligned}
\end{equation}
\begin{equation}
    \frac{d \Delta \Phi}{d t} = -\frac{1}{24} r_e \kappa^4 \left< \gamma \right> \left(\left| V \right|^2 + \left| U \right|^2 \right) \sin 2 \Delta \Phi
    + \frac{1}{8} \kappa \left[\frac{1}{4} \lambda \beta_{z0}  
    - \kappa^2 \left(1- 2\lambda \right) \right] \left<\gamma \right>^{-\frac{3}{2}} \left(\left| V \right|^2 - \left| U \right|^2 \right) \sin^2 \Delta \Phi. \label{eq:dot_DPhi}
\end{equation}
\end{widetext}
Note when doing exchange of $U$ and $V$ for the $y$ direction, one also has to change the $\pm$ sign of $\Delta \Phi$.

To further simplify we notice Eq.~(\ref{eq:dot_U_ave_2}) can be rewritten with the help of Eq.~(\ref{eq:dot_gamma_ave})
\begin{widetext}
\begin{equation}
    \frac{d{\left| U \right|}}{d t} = -\frac{1}{4} \frac{\left<\dot{\gamma} \right>}{\left< \gamma \right>} \left| U \right| - \frac{1}{8} r_e \kappa^4 \left< \gamma \right> \left[\left| U \right|^3 + \frac{4 - \cos 2 \Delta \Phi}{3}\left| V \right|^2 \left| U \right| \right] - \frac{1}{16} \kappa \left[\frac{1}{4} \lambda \beta_{z0} - \kappa^2 \left(1-2\lambda\right)\right] \left< \gamma \right>^{-\frac{3}{2}} \left| V \right|^2 \left| U \right| \sin 2 \Delta \Phi,
\end{equation}
\end{widetext}
which reproduces Eq.~(66) in Ref.~\cite{MZengNJP2021} if $V=0$. Introduce
\begin{eqnarray}
    S_{x} &=& \kappa \left< \gamma \right>^{\frac{1}{2}} \left| U \right|^2, \\
    S_{y} &=& \kappa \left< \gamma \right>^{\frac{1}{2}} \left| V \right|^2,
\end{eqnarray}
which have the physical meaning of the areas (divided by $2\pi$) of the ellipses encircled by the particle trajectory in $x$-$p_x$ and $y$-$p_y$ phase spaces. Then Eqs.~(\ref{eq:dot_gamma_ave}), (\ref{eq:dot_zeta_ave}), (\ref{eq:dot_U_ave_2}), (\ref{eq:dot_Phix}) and (\ref{eq:dot_DPhi}) can be rewritten as
\begin{widetext}
\begin{equation}
    \left<\dot{\gamma} \right> = - E_{z0} \beta_{z0} - \frac{1}{3} r_e \kappa^3 \left< \gamma \right>^{\frac{3}{2}} \left(S_x + S_y \right), \label{eq:dot_gamma_ave_withS}
\end{equation}
\begin{equation}
    \dot{\left<\zeta \right>} = \frac{1}{2} \gamma_w^{-2} - \frac{1}{2} \left<\gamma\right>^{-2} - \frac{1}{4} \kappa \left< \gamma \right>^{-\frac{3}{2}} \left(S_x + S_y \right), \label{eq:dot_zeta_ave_withS}
\end{equation}
\begin{equation}
    \dot{S}_x = -\frac{1}{4} r_e \kappa^3 \left< \gamma \right>^{\frac{1}{2}} \left(S_x^2 + \frac{4 - \cos 2 \Delta \Phi}{3} S_x S_y \right)
    - \frac{1}{8} \left[\frac{1}{4} \lambda \beta_{z0} - \kappa^2 \left(1-2\lambda\right)\right] \left< \gamma \right>^{-2} S_x S_y \sin 2 \Delta \Phi, \label{eq:dot_Sx_withS}
\end{equation}
\begin{equation}
\begin{aligned}
    \dot{\Phi}_x =& \frac{1}{24} r_e \kappa^3 \left< \gamma \right>^{\frac{1}{2}} S_y \sin 2\Delta\Phi + \frac{1}{64} \lambda \beta_{z0} \left< \gamma \right>^{-2} \left(S_x + S_y \cos 2\Delta \Phi \right) \\
    &- \frac{1}{16} \kappa^2 \left< \gamma \right>^{-2} \left[S_x + 2\lambda S_y + \left(1-2\lambda\right) S_y\cos 2\Delta \Phi\right] - \frac{1}{4} \kappa \lambda \left< \gamma \right>^{-\frac{5}{2}}, \label{eq:dot_Phix_withS}
\end{aligned}
\end{equation}
\begin{equation}
    \frac{d \Delta \Phi}{d t} = - \frac{1}{24} r_e \kappa^3 \left< \gamma \right>^{\frac{1}{2}} \left(S_y + S_x \right) \sin 2 \Delta \Phi
    + \frac{1}{8} \left[\frac{1}{4}\lambda \beta_{z0} - \kappa^2 \left(1- 2\lambda \right)\right] \left< \gamma \right>^{-2} \left(S_y - S_x \right) \sin^2 \Delta \Phi. \label{eq:dot_DPhi_withS}
\end{equation}
\end{widetext}
It is generally safe to take $\beta_{z0}=1$ here. But Eq.~(\ref{eq:betaz0}), or $\beta_{z0} = 1 - \frac{1}{2}\left[\left<\gamma\right>^{-2} + \frac{1}{2}\kappa\left<\gamma\right>^{-3/2} \left(S_x+S_y\right)\right]$, gives a better accuracy. The above long-term equations, Eqs.~(\ref{eq:dot_gamma_ave_withS}) - (\ref{eq:dot_DPhi_withS}), show that the BO experiences acceleration (for $E_{z0}<0$) or deceleration (for $E_{z0}>0$), radiation damping, longitudinal phase drift, and betatron phase shift. These equations may be used for the long-term behavior of BO without resolving the betatron period.

\section{\label{sec:two_regimes}Discussion on two regimes}
From Eqs.~(\ref{eq:dot_Sx_withS}) - (\ref{eq:dot_DPhi_withS}) we note two regimes. One is the RR dominant regime, where $r_e \left<\gamma\right>^{5/2}\gg 1$, so that the first terms in Eqs.~(\ref{eq:dot_Sx_withS}) - (\ref{eq:dot_DPhi_withS}) dominate. This regime has been discussed before~\cite{IYKostyukovPRSTAB2012}, although only for the linearly polarized case $\Delta \Phi = 0$ (so that the ratio between $S_x$ and $S_y$ is a constant). The other is the betatron phase shift dominant regime, where $r_e \left<\gamma\right>^{5/2}\ll 1$, so that the remaining terms in Eqs.~(\ref{eq:dot_Sx_withS}) - (\ref{eq:dot_DPhi_withS}) dominate. These terms were previously proposed~\cite{MZengNJP2021}, but the betatron phase shift is found for the first time in the present work.

In the RR dominant regime, an interesting phenomenon is that an elliptical polarization (in the $x$-$y$ plane) always approaches linear polarization, because $\Delta \Phi$ always approaches the nearest integer multiple of $\pi$ according to Eq.~(\ref{eq:dot_DPhi_withS}). This phenomenon can also be viewed by rotating the $x$ axis to the major axis of the ellipse, so that $S_x>S_y$ and $\Delta \Phi = \pi/2$. Define $R=S_y/S_x$ and perform time derivative with the help of Eq.~(\ref{eq:dot_Sx_withS})
\begin{equation}
    \dot{R} = -\frac{1}{6} r_e \kappa^3 \left<\gamma\right>^{\frac{1}{2}} R \left(S_x - S_y\right) <0, \label{eq:dot_R}
\end{equation}
which suggests that the ellipse monotonically becomes thinner.

The betatron phase shift dominant regime requires a moderate $\gamma$, or straightforwardly $r_e\rightarrow 0$, which corresponds to very dilute plasma case, leads to a constant $S \equiv S_x+S_y$. It can be proved that the time integral of Eq.~(\ref{eq:dot_zeta_ave_withS}) reproduces Eq.~(6) in Ref.~\cite{AFerraPousaSR2019}, which is the $\left<\zeta\right>$ drift, in the case that $\left<\gamma\right>$ linearly depends on $t$. In another case that $\left<\zeta\right>$ drifts around the zero point of $E_{z0}$, the drift frequency can be obtained by using Eqs.~(\ref{eq:dot_gamma_ave_withS}) and (\ref{eq:dot_zeta_ave_withS}), and asserting $E_{z0}=\lambda \left<\zeta\right>$
\begin{equation}
    \omega_{\left<\zeta\right>} = \sqrt{\lambda \beta_{z0} \left(1 + \frac{3}{8} \kappa \left< \gamma \right>^{\frac{1}{2}} S \right) \left< \gamma \right>^{-3}}. \label{eq:omega_zeta}
\end{equation}

We also note the angular momentum $L_z = \gamma x \beta_y - \gamma y \beta_x$ and its changing rate
\begin{eqnarray}
    \left<L_z\right> &=& - S_x^{\frac{1}{2}} S_y^{\frac{1}{2}} \sin \Delta \Phi, \\
    \dot{\left<L_z \right>} &=& -\frac{1}{3} r_e \kappa^3\left< \gamma \right>^{\frac{1}{2}} \left(S_x + S_y \right) \left<L_z\right>,
\end{eqnarray}
which obeys the law of conservation of angular momentum if $\left<L_z\right>=0$ initially, or $r_e\rightarrow 0$. Especially for $r_e\rightarrow 0$, the particle trajectory in the $x$-$y$ plane generally encircles an ellipse with constant area and shape, which also has precession leading to the rotation of the major and minor axes of the ellipse.

\section{\label{sec:num}Numerical comparison of long-term equations and the original ones}
\begin{figure}
    \centering
    \begin{overpic}[width=0.48\textwidth]{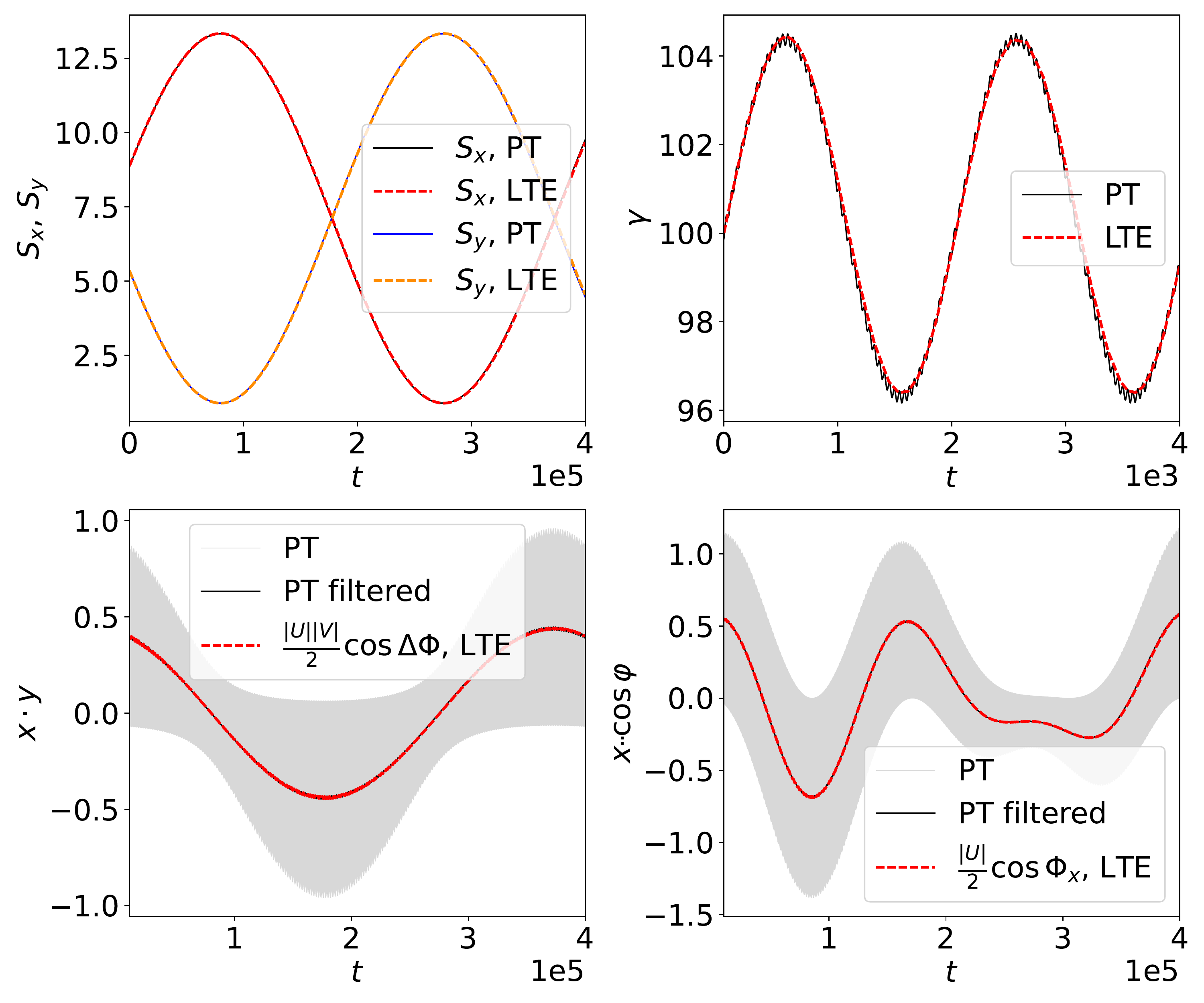}
        \put(16,54){(a)}
        \put(65,54){(b)}
        \put(15,12){(c)}
        \put(65,35){(d)}
    \end{overpic}
    \caption{\label{fig:change_zeta0}The numerical comparison of the LTEs and the original equations solved by PTracker in the betatron phase shift dominant regime. (a) $S_x$ and $S_y$ change with time, but $S_x+S_y$ is a constant. (b) $\gamma$ has oscillation frequencies of $2\omega_{\beta}\approx 0.14$ due to the BO and $\omega_{\left<\zeta\right>}\approx 3.11 \times 10^{-3}$ due to the drift oscillation of $\left<\zeta\right>$. (c) The gray curve shows $x\cdot y$ obtained from PT, and the black curve shows its low-pass filtered result, which is compared with the LTE solution according to Eq.~(\ref{eq:xy_treat}). (d) The gray curve shows $x\cdot \cos \varphi$ obtained from PT, and the black curve shows its low-pass filtered result, which is compared with the LTE solution according to Eq.~(\ref{eq:x_treat}).}
\end{figure}

\begin{figure}
    \centering
    \begin{overpic}[width=0.48\textwidth]{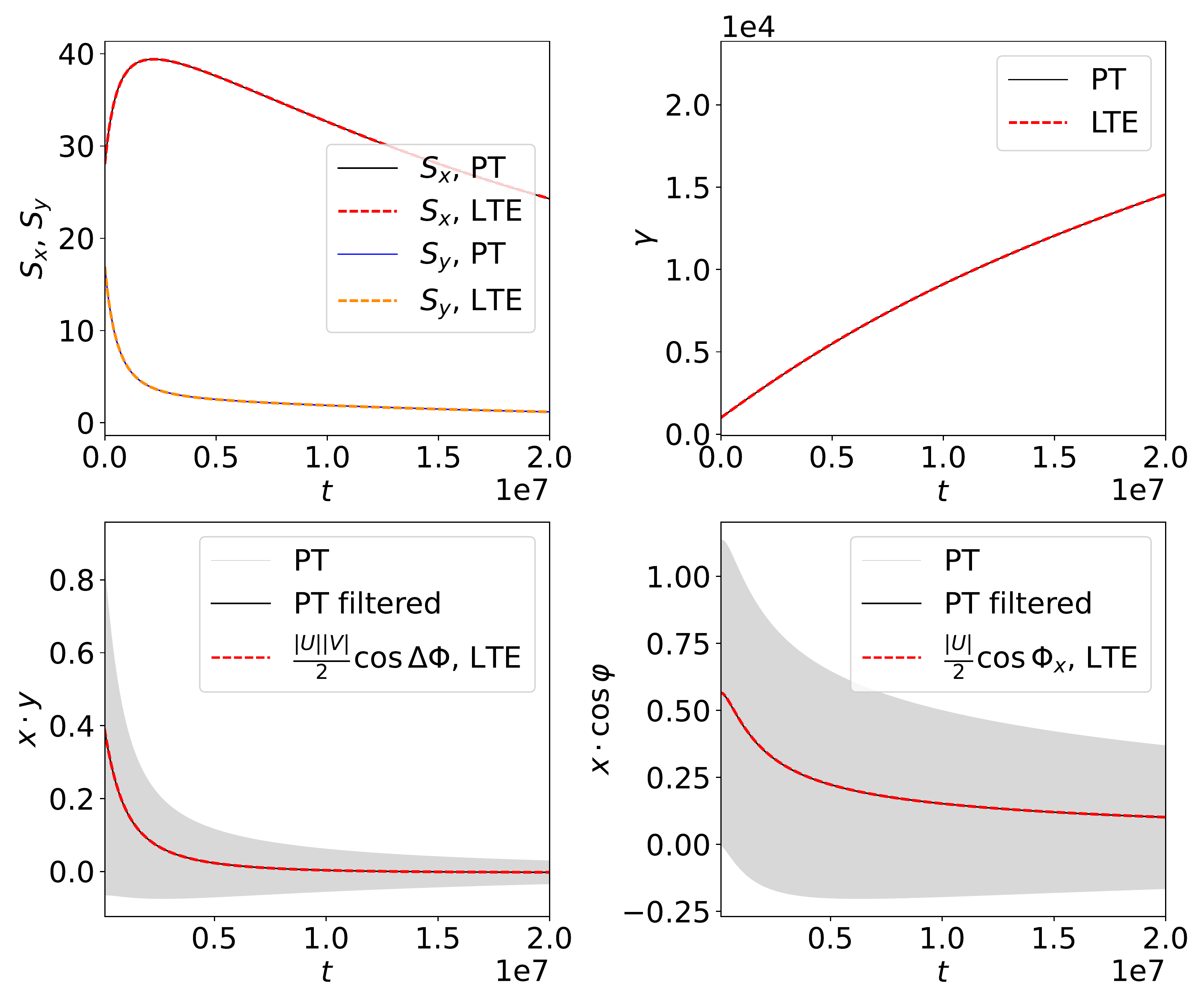}
        \put(12,54){(a)}
        \put(64,74){(b)}
        \put(10,35){(c)}
        \put(63,35){(d)}
    \end{overpic}\caption{\label{fig:change_fz0_plusRRforce}The numerical comparison of the LTEs and the original equations solved by PTracker in the transition between the betatron phase shift dominant and the RR dominant regimes. (a) $S_x+S_y$ is initially approximately a constant, but decreases later. (b) $\gamma$ increases due to the acceleration field, and passes the regime transition at $\gamma=r_e^{-2/5} = 10^4$. (c) and (d) show the same treatments as in Fig.~\ref{fig:change_zeta0} (c) and (d).}
\end{figure}

To verify the LTEs, we solve them numerically using the backward-differentiation formulas (BDF) in the SciPy integration package~\cite{BDF}. Meanwhile, the original equations of motion with the force expressions Eqs.~(\ref{eq:fz}) - (\ref{eq:fy}) are solved by Runge-Kutta 4th order method using the code PTracker (PT)~\cite{PTracker}. We choose four cases with their parameters and initial values listed in Tab.~\ref{tab:para}, and the comparison results are plotted from Fig.~\ref{fig:change_zeta0} to \ref{fig:change_DeltPhi_plusRR}. Note that $\Phi_x$ cannot be obtained directly from PT. Thus we perform the following treatment to the PT results
\begin{equation}
    x \cdot \cos \varphi = \frac{\left|U\right|}{2} \left[\cos \Phi_x + \cos \left(2 \varphi + \Phi_x \right) \right], \label{eq:x_treat}
\end{equation}
because $x = \left|U\right| \cos \left(\varphi + \Phi_x \right)$, where $\varphi$ is obtained by numerical integral based on Eq.~(\ref{eq:varphi}). Then $\cos \Phi_x$ can be obtained by a low-pass filter. Similar treatment is performed to obtain $\cos \Delta \Phi$, according to
\begin{equation}
    x \cdot y = \frac{\left|U\right| \left|V\right|}{2} \left[\cos \Delta \Phi + \cos \left(2 \varphi + \Phi_x + \Phi_y \right) \right]. \label{eq:xy_treat}
\end{equation}

A case in the betatron phase shift dominant regime is shown in Fig.~\ref{fig:change_zeta0}. We see $S_x+S_y$ is a constant, although $S_x$, $S_y$ and $\Delta \Phi$ change gradually. The approximate ``phase-locking'' is chosen, i.e.\ $\gamma_w \approx \gamma_{z0}$, thus $\left<\zeta\right>$ oscillates near the zero point of $E_z$ with the drift frequency $\omega_{\left<\zeta\right>}$ according to Eq.~(\ref{eq:omega_zeta}). $\left<\gamma\right>$ oscillates with the same drift frequency as shown in Fig.~\ref{fig:change_zeta0} (b).

A second case during the transition of the two regimes is shown in Fig.~\ref{fig:change_fz0_plusRRforce}. $S_x+S_y$ is approximately a constant initially, and starts to decrease near the regime transition $\gamma=r_e^{-2/5}$.

A third case in the RR dominant regime is shown in Fig.~\ref{fig:change_R_plusRR}. The initial values are chosen so that the particle trajectory in the $x$-$y$ plane is a ellipse with its major axis laying on the $x$ axis. As shown in Fig.~\ref{fig:change_R_plusRR} (a), $R=S_y/S_x$ decreases monotonically, as predicted by Eq.~(\ref{eq:dot_R}).

The last case shown in Fig.~\ref{fig:change_DeltPhi_plusRR} is also in the RR dominant regime, but the particle trajectory in the $x$-$y$ plane is a oblique ellipse. As shown in Fig.~\ref{fig:change_DeltPhi_plusRR} (c), $\Delta \Phi$ gradually approaches $\pi$, which is in accordance with the discussion in Sec.~\ref{sec:two_regimes}.

In all these plots, the results from PT and LTEs show agreement with high accuracy, demonstrating the correctness of LTEs. Because the BO frequency is the highest frequency in our physical process, the LTEs largely reduce the numerical complexity and meanwhile keep the long-term accuracy.

\begin{figure}
    \centering
    \begin{overpic}[width=0.48\textwidth]{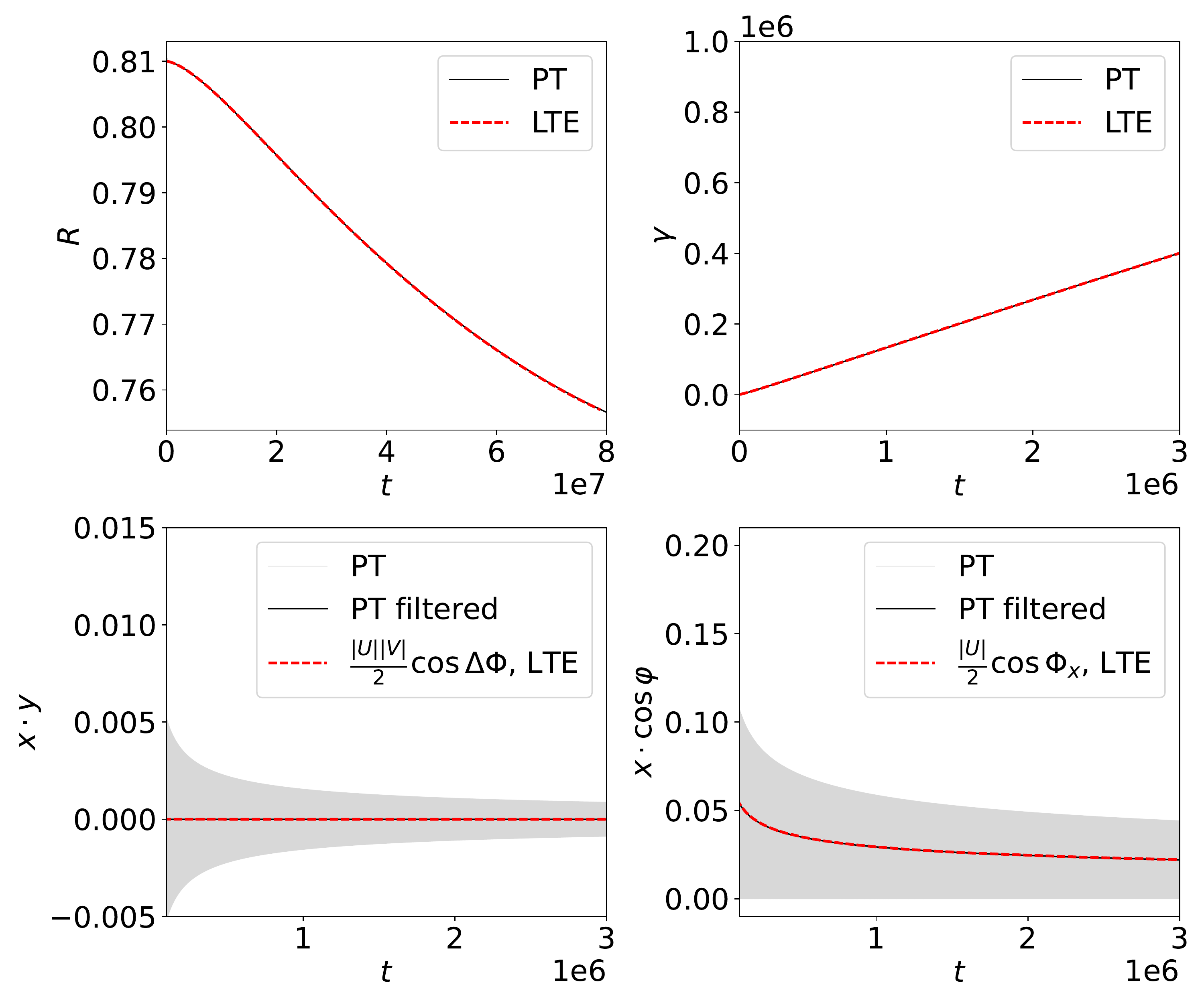}
        \put(18,53){(a)}
        \put(65,74){(b)}
        \put(15,34){(c)}
        \put(65,34){(d)}
    \end{overpic}\caption{\label{fig:change_R_plusRR}The numerical comparison of the LTEs and the original equations solved by PTracker in the RR dominant regime. $\Delta \Phi = \pi/2$ and $S_x>S_y$, thus the major axis of the particle trajectory ellipse lays on the $x$ axis. (a) $R=S_y/S_x$ decreases with time due to Eq.~(\ref{eq:dot_R}), thus the ellipse is getting thinner. (b) $\gamma$ increases due to the acceleration field. (c) and (d) show the same treatments as in Fig.~\ref{fig:change_zeta0} (c) and (d). }
\end{figure}
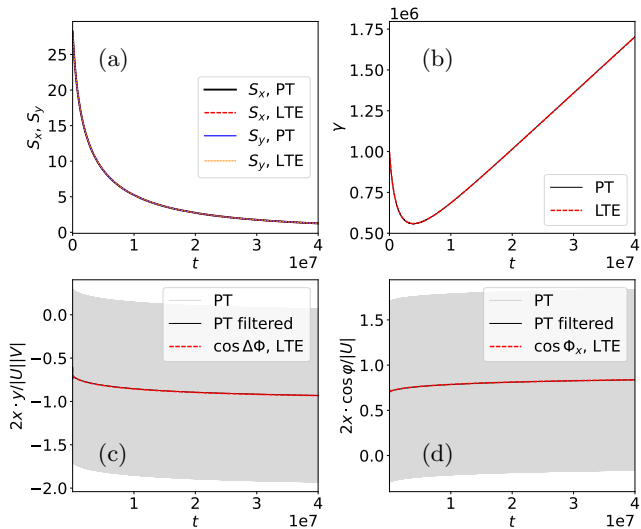
\begin{figure}
    \centering
    \begin{overpic}[width=0.48\textwidth]{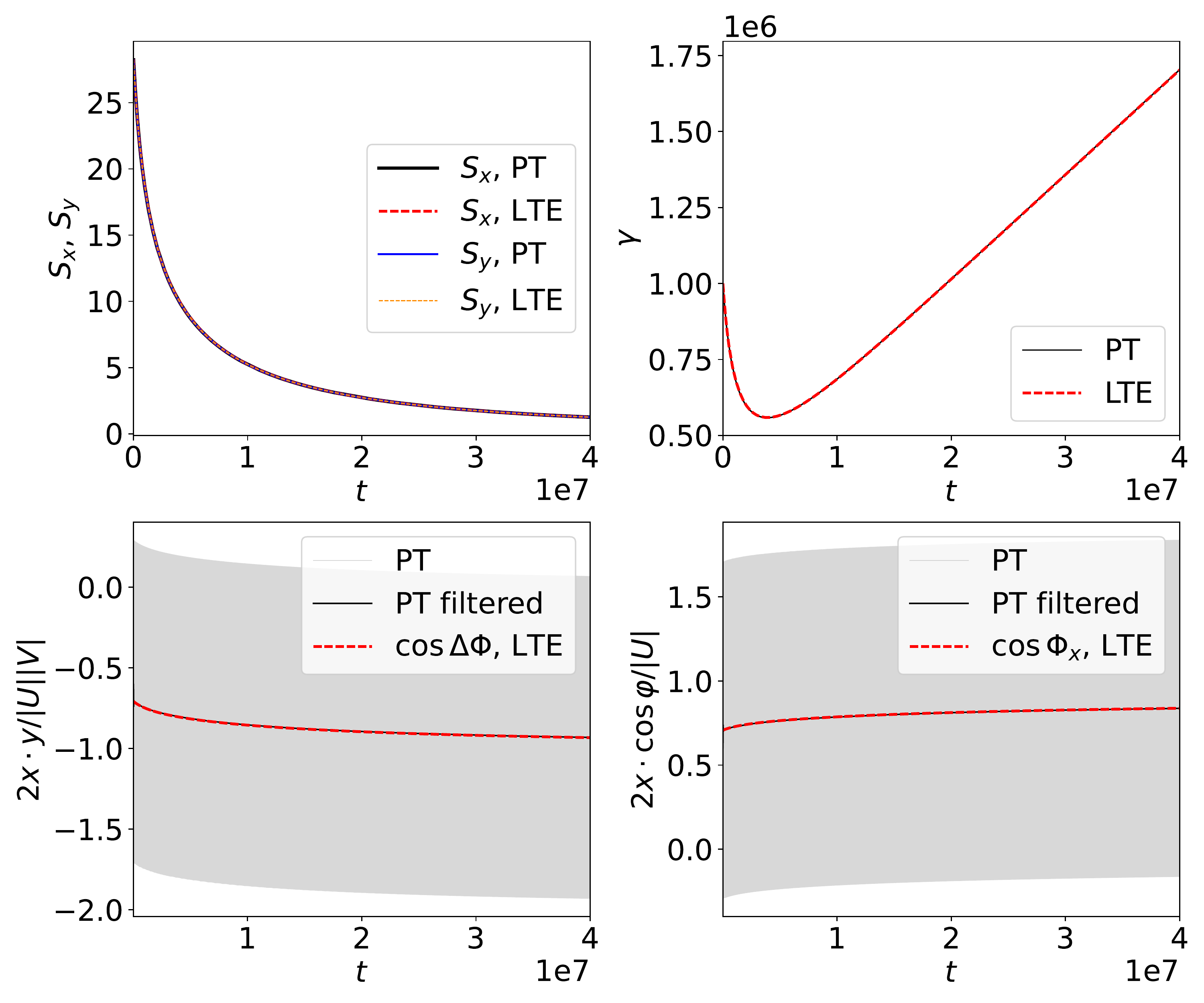}
        \put(15,73){(a)}
        \put(65,73){(b)}
        \put(15,12){(c)}
        \put(65,12){(d)}
    \end{overpic}\caption{\label{fig:change_DeltPhi_plusRR}The numerical comparison of the LTEs and the original equations solved by PTracker in the RR dominant regime. $S_x=S_y$, thus the particle trajectory in the $x$-$y$ plane is an oblique ellipse. (a) $S_x$ and $S_y$ decrease with the same rate. (b) $\gamma$ initially decreases with time because the longitudinal RR damping is stronger than the acceleration. Later the RR damping becomes weaker due to the decrease of $S_x$ and $S_y$, and $\gamma$ increases. (c) and (d) show the same treatments as in Fig.~\ref{fig:change_zeta0} (c) and (d), but the oscillation amplitudes are divided so that the changes of $\Delta \Phi$ and $\Phi_x$ are clearer. $\Delta \Phi$ is between $\pi/2$ and $\pi$, thus $\Delta \Phi$ gradually approaches $\pi$.}
\end{figure}
\begin{table}
  \begin{center}
  \caption{\label{tab:para} The cases for comparing PT and LTEs}
    \begin{tabular}{|c|c|c|c|c|c|c|c|c|c|c|c|} 
    \hline
      \multirow{2}*{\textbf{Case}} & \multicolumn{5}{c|}{\textbf{Parameters}} & \multicolumn{6}{c|}{\textbf{Initial Values}} \\ 
      \cline{2-12}
      ~ & $E_z$ & $\lambda$ & $\kappa$ & $r_{e}$ & $\gamma_w$ & $\left|U\right|$ & $\left|V\right|$ & $\Phi_x$ &  $\Phi_y$ & $\left<\gamma \right>$ & $\left<\zeta \right>$ \\
       \hline
       Fig.~\ref{fig:change_zeta0} & $\lambda \zeta$ & $\dfrac{1}{4}$ & $\dfrac{1}{\sqrt{2}}$ & 0 & 14 & 1.12 & 0.87 & 0 & $\dfrac{\pi}{6}$ & $10^2$ & -0.05 \\
       \hline
       Fig.~\ref{fig:change_fz0_plusRRforce} & -0.001 & 0 & $\dfrac{1}{\sqrt{2}}$ & $10^{-10}$ & $10^4$ & 1.12 & 0.87 & 0 & $\dfrac{\pi}{6}$ & $10^3$ & 0 \\
       \hline
       Fig.~\ref{fig:change_R_plusRR} & $\lambda \zeta$ & $\dfrac{1}{2}$ & $\dfrac{1}{\sqrt{2}}$ & $10^{-10}$ & $10^4$ & 0.2 & 0.18 & 0 & $\dfrac{\pi}{2}$ & $10^3$ & -0.1 \\
       \hline
       Fig.~\ref{fig:change_DeltPhi_plusRR} & -0.1 & 0 & $\dfrac{1}{\sqrt{2}}$ & $10^{-10}$ & $10^4$ & 0.2 & 0.2 & $\dfrac{\pi}{4}$ & $\pi$ & $10^6$ & 0 \\
       \hline
    \end{tabular}
  \end{center}
\end{table}

\section{Conclusions}
We have established a three-dimensional betatron oscillation model including radiation reaction to study the long-term behavior of an electron in laser or beam driven plasma wakefield. The original equations of motion have been expressed by the transverse oscillation terms as Eqs.~(\ref{eq:dot_gamma_rr}) - (\ref{eq:dot_py_rr}), and then averaged in one betatron period to obtain the long-term equations Eqs.~(\ref{eq:dot_gamma_ave_withS}) - (\ref{eq:dot_DPhi_withS}). The conditions of our model are $r^2\ll \gamma$, $r^2\gamma \gg 1$ and $r \gamma r_e /2\alpha \ll 1$, as discussed in Appx.~\ref{app:Eq_motion}. Our model, on one hand, reproduces previous results such as longitudinal deceleration and transverse damping, and on the other hand reveals new phenomena such as longitudinal phase drift oscillation, betatron phase shift and betatron polarization change. Two regimes with distinct behaviors, determined by $r_e\gamma^{5/2}$, are discussed in Sec.~\ref{sec:two_regimes}, and are demonstrated by numerical methods in Sec.~\ref{sec:num}. The numerical comparisons of the long-term equations and the original equations of motion show the high accuracy of our model. This model can be fundamental for future plasma based high-energy accelerators and colliders~\cite{WLeemansPT2009}.

\begin{acknowledgments}
MZ greatly appreciates the fruitful discussion on the averaging method with Igor Kostyukov and Anton Golovanov from Institute of Applied Physics RAS, Russia. This work is supported by Research Foundation of Institute of High Energy Physics, Chinese Academy of Sciences (Grant Nos.\ E05153U1, E15453U2).
\end{acknowledgments}

\appendix
\section{\label{app:norm}Plasma normalization units}
The plasma normalization units are used throughout the paper, as listed in Tab.~\ref{tab:norm}, where $c$ is the speed of light in vacuum, $\omega_p$ is the plasma frequency, $e$ is the elementary charge, and $m_e$ is the electron mass. For example, the time is normalized to $\omega_p^{-1}$, means any time related quantity such as $t$ in this paper actually means $\omega_p t$ in the unnormalized form.
\begin{table}[h!]
  \begin{center}
    \caption{\label{tab:norm} The plasma normalization units}
    \begin{tabular}{|c|c|c|} 
    \hline
      \textbf{Physical quantities} & \textbf{Variables} & \textbf{Normalization units}\\
      \hline
      time & $t$ & $\omega_p^{-1}$\\
      frequency & $\omega$ & $\omega_p$\\
      length & $x, y, z, r_e$ & $c/\omega_p$\\
      velocity & $v$ & $c$\\
      momentum & $p$ & $m_e c$\\
      angular momentum & $L$ & $m_e c^2/ \omega_p$\\
      electric field & $E$ & $m_e c \omega_p/e$\\
      magnetic field & $B$ & $m_e \omega_p/e$ (in SI)\\
      force & $f$ & $m_e c \omega_p$\\
      \hline
    \end{tabular}
  \end{center}
\end{table}

\section{\label{app:rules}Symbols and rules}
If any variable $X$, either real or complex, can be expressed as $X = \left<X\right> + X_1$, where $\left<\right>$ means taking average in the betatron period time scale, and $X_1$ is the BO term, taking average and derivative can permute
\begin{equation}
    \left<\frac{d}{dt}X\right> = \frac{d}{dt}\left<X\right>.
\end{equation}
We use a dot on the top to express the time derivative if there is no ambiguity. We have the order-of-magnitude estimation
\begin{equation}
    \dot{X_1}\sim \omega_{\beta} X_1 \sim \gamma^{-\frac{1}{2}} X_1. \label{dot_estim}
\end{equation}
If $\left|X_1\right| \ll \left|\left<X\right>\right|$, for any power $\alpha$ we have
\begin{equation}
    \left<X^\alpha\right> = \left<X\right>^\alpha \left[1+\mathcal{O}\left(\frac{X_1^2}{\left<X\right>^2}\right)\right]. \label{eq:rule3}
\end{equation}
And if another variable $Y=\left<Y\right> + Y_1$ also has $\left|Y_1\right| \ll \left|\left<Y\right>\right|$,
\begin{equation}
    \left<XY\right> = \left<X\right>\left<Y\right> \left[1+\mathcal{O}\left(\frac{X_1 Y_1}{\left<X\right>\left<Y\right>}\right)\right].
\end{equation}
If $X$ is a complex, taking average and modulus can permute
\begin{equation}
    \left<\left|X\right|\right>=\left|\left<X\right>\right|\left[1+\mathcal{O}\left(\frac{\left|X_1\right|^2}{\left|\left<X\right>\right|^2}\right)\right].
\end{equation}
However, taking modulus and derivative cannot permute.

\section{\label{app:Eq_motion}Equations of motion expressed by transverse oscillations}
In Eqs.~(\ref{eq:fz}) - (\ref{eq:fy}), the longitudinal and transverse oscillations are coupled. As shown in the following, the longitudinal variables $\zeta_1$ and $\beta_z$ are dependent variables which can be expressed by the transverse ones.

We treat $\vec{f}^{\rm rad}$ as a perturbation and omit it first. On one hand we have
\begin{equation}
\begin{aligned}
    \gamma^{-2} &= 1 - \beta_z^{2} - \beta_x^{2} - \beta_y^{2} \\
    &= \gamma_{z0}^{-2} - 2\beta_{z0}\dot{\zeta_1} - \beta_x^{2} - \beta_y^{2} + \mathcal{O}\left(\dot{\zeta_1}^2\right),
\end{aligned}
\end{equation}
where $\gamma_{z0} = \left(1-\beta_{z0}^2\right)^{-1/2}$. By taking average we get
\begin{equation}
    \gamma_{z0}^{-2} \approx \left<\gamma\right>^{-2} + \left<\beta_x^2\right> + \left<\beta_y^2\right>. \label{eq:gammaz0}
\end{equation}
Write $\gamma = \left<\gamma\right> + \gamma_1$ in the form
\begin{equation}
    \gamma^{-2} = \left<\gamma\right>^{-2}\left[1 - 2\frac{\gamma_1}{\left<\gamma\right>} + \mathcal{O}\left(\frac{\gamma_1^2}{\left<\gamma\right>^2}\right)\right],
\end{equation}
we have
\begin{equation}
    \gamma_1 \approx \left[\frac{\beta_x^2 - \left<\beta_x^2\right>}{2} + \frac{\beta_y^2 - \left<\beta_y^2\right>}{2} + \beta_{z0} \dot{\zeta_1} \right]\left<\gamma \right>^3. \label{eq:gamma1}
\end{equation}
On the other hand,
\begin{equation}
    \dot{\gamma} = -E_{z0} \beta_{z0} - \lambda\beta_{z0}\zeta_1 - \kappa^2\left(1-\lambda\right)\left(x\beta_x + y\beta_y\right)
\end{equation}
by applying $\dot{\gamma} = -\vec{\beta}\cdot \vec{E}$ and Eqs.~(\ref{eq:Ez}) and (\ref{eq:Er}), or
\begin{equation}
    \dot{\gamma_1} = -\lambda\beta_{z0}\zeta_1 - \kappa^2\left(1-\lambda\right)\left(x\beta_x + y\beta_y\right). \label{eq:dot_gamma1}
\end{equation}
Note Eq.~(\ref{dot_estim}), Eq.~(\ref{eq:gamma1}) seams incompatible with Eq.~(\ref{eq:dot_gamma1}), unless
\begin{equation}
    \dot{\zeta_1} = - \frac{\beta_x^2- \left<\beta_x^2\right>}{2} - \frac{\beta_y^2- \left<\beta_y^2\right>}{2},
\end{equation}
which leads to
\begin{equation}
    \zeta_1 = - \frac{x\beta_x + y\beta_y}{4},
\end{equation}
which is a general form of Eq.~(18) in Ref.~\cite{MZengNJP2021}. Then
\begin{equation}
    1-\beta_z = 1-\beta_{z0}-\dot{\zeta_1} = \frac{1}{2}\left(\left<\gamma\right>^{-2} + \beta_x^2 + \beta_y^2\right),
\end{equation}
and the formulas of 3D BO with negligible RR are
\begin{align}
    \dot{\gamma} &= -E_{z0}\beta_{z0} + \left(\frac{\lambda\beta_{z0}}{4}+\kappa^2\lambda-\kappa^2\right) \left(x \beta_x + y \beta_y\right), \label{eq:dot_gamma}\\
    \dot{p_z} &= -E_{z0} + \lambda\left(\frac{1}{4} + \kappa^2\right) \left(x \beta_x + y \beta_y\right),\\
    \dot{p_x} &= -\kappa^2 x + \frac{\kappa^2 \lambda}{2}\left(\left<\gamma \right>^{-2} + \beta_x^2 + \beta_y^2 \right)x, \label{eq:dot_px}\\
    \dot{p_y} &= -\kappa^2 y + \frac{\kappa^2 \lambda}{2}\left(\left<\gamma \right>^{-2} + \beta_x^2 + \beta_y^2 \right)y. \label{eq:dot_py}
\end{align}
From Eq.~(\ref{eq:dot_gamma}) we may write
\begin{equation}
    \gamma = \left<\gamma\right> + \left(\frac{\lambda\beta_{z0}}{4}+\kappa^2\lambda-\kappa^2\right) \frac{x^2 -\left<x^2\right> + y^2 - \left<y^2\right>}{2},
\end{equation}
indicating the prerequisite of the above derivation, which has used Eq.~(\ref{eq:rule3}), is $r^2\ll \left<\gamma\right>$.

Now we consider RR as a perturbation. The LAD equation for the RR four-force is~\cite{Dirac1938}
\begin{equation}
    F_{\mu}^{\rm rad} = \frac{2}{3}r_e \left[\frac{d^2 P_\mu}{d\tau^2} + \left( \frac{d P_\nu}{d\tau} \frac{d P^{\nu}}{d\tau} \right)P_\mu \right]. 
    \label{eq:frad}
\end{equation}
with the metric $\left(1, -1, -1, -1 \right)$, where $r_e$ is the classical electron radius (also normalized to $k_p^{-1}$), $P_\mu$ is the four-momentum, and $\tau$ is the proper time ($d\tau=dt/\gamma$). Use Eqs.~(\ref{eq:dot_gamma}) - (\ref{eq:dot_py}) we can verify
\begin{equation}
    \frac{d P_\nu}{d\tau} \frac{d P^{\nu}}{d\tau} = \gamma^2\left(\dot{\gamma}^2 - \left| \dot{\vec{p}}\right|^2\right) \approx -\gamma^2\left(\dot{p_x}^2 + \dot{p_y}^2 \right)
\end{equation}
as long as $r^2\gamma^2 \gg 1$. We can also prove that the first term in Eq.~(\ref{eq:frad}) is negligible compared with the second term as long as $r^2\gamma \gg 1$. Finally the equations of motion expressed by the transverse oscillations are obtained as Eqs.~(\ref{eq:dot_gamma_rr}) - (\ref{eq:dot_py_rr}). As it has been discussed in Ref.~\cite{MZengNJP2021} and \cite{GolovanovNJP2022}, this classical RR model is valid as long as $r \gamma r_e /2\alpha \ll 1$, where $\alpha$ is the fine structure constant.
\bibliography{main}
\end{document}